\documentclass[aps,prc,twocolumn,groupedaddress,showpacs,fleqn]{revtex4}

\usepackage{graphicx}
\usepackage{bm}
\usepackage{mathrsfs}
\begin{document}
\title{Luneburg-lens-like universal structural Pauli attraction 
in   nucleus-nucleus interactions: 
origin of emergence of   cluster structures and nuclear rainbows
 }

\author{S. Ohkubo }
\affiliation {Research Center for Nuclear Physics, Osaka University, 
Ibaraki, Osaka 567-0047, Japan }

\date{\today}

\begin{abstract}
 The Pauli exclusion principle plays an important role in many-body fermion systems
 preventing them from collapsing by repulsion. For example, the Pauli principle causes a repulsive
 potential at short distances between two $\alpha$ particles. On the other hand, the  existence
  of nuclear  rainbows demonstrates that the inter-nuclear potential is sufficiently  attractive in
 the internal region to cause  refraction. The two concepts of repulsion and attraction are
 seemingly irreconcilable. Contrary to traditional understanding, it is shown that the Pauli principle
 causes a  {\it universal structural Pauli attraction}  between  nuclei rather than a
   {\it structural repulsive core}. Through systematic studies of  $\alpha$+$\alpha$, $\alpha$+$^{16}$O, 
$\alpha$+$^{40}$Ca  and $^{16}$O+$^{16}$O systems, it is shown that the emergence of 
 cluster structures near the  threshold energy at low energies and nuclear rainbows
 at high energies is a direct consequence of the Pauli principle.
  
\end{abstract}

\maketitle

\par
Why do  cluster structures appear near the threshold energy, while molecular
 resonances occur at  higher excitation energies, and nuclear rainbows at even high energies? 
 The threshold rule, molecular resonance theory and nuclear rainbow theory
 have been   proposed 
  and  extensively studied for more than fifty years.
Until now, these independent theories - each describing successfully different facets
 of nuclear structure - had not been thought to be closely connected in the level of 
 the fundamental  principle.
I will show that  they share a common  {\it raison d'\^etre}: They are all   a direct
 consequence of the universal Pauli attraction.

\par
The Pauli exclusion plays an important role in nuclei.
The shell structure, 
in which nucleons behave like  independent particles  in a mean field potential
and  persist throughout the periodic table, is 
a   consequence of the Pauli principle  and  the short-range character of the  nuclear force
 \cite{Bohr1969}.
The Pauli principle also plays an important  role between   nuclei 
 at small distances where they overlap.
 For the typical $\alpha$+$\alpha$ system,   microscopic  Resonationg Group Method (RGM) studies
 have revealed \cite{Tamagaki1962,Tamagaki1965,Tamagaki1968,Hiura1972} that 
the inter-nuclear interaction for $S$  and $D$ waves has a   {\it repulsive}   core
 at short   distances
and angular-momentum ($L$)-dependent  {\it shallow}   attraction in the outer region.
   The repulsive core
explains the experimental phase  shifts in $\alpha$+$\alpha$ scattering and the 
  well-developed $\alpha$ cluster structure of $^8$Be well
\cite{Tamagaki1965,Tamagaki1968,Hiura1972,Ali1966}. 
The repulsive core was found to be a potential
 representation of the damped inner oscillations of the inter-cluster wave functions, 
with  the energy-independent node at around 2 fm caused by the Pauli principle 
\cite{Tamagaki1968,Hiura1972}.
This is  known as the {\it structural  repulsive core} \cite{Otsuki1965}.
  For heavy ion systems such as $^{16}$O+$^{16}$O, the existence of a  repulsive core
 at short distances has also been  shown in   microscopic  model calculations
\cite{Tohsaki1975,Ando1978,Buck1977}.

 \par
Although the so-called standard optical potential model with a Woods-Saxon
 form factor witnessed tremendous success in the studies of 
light-ion and heavy-ion scattering and reactions  \cite{Hodgson1978},
 it could not describe   the Backward Angle Anomaly (BAA) or Anomalous Large Angle
 Scattering (ALAS) in $\alpha$+$^{16}$O and  $\alpha$+$^{40}$Ca scattering
 \cite{Brink1985}.
 This was shown to be resolved
  using a {\it non-standard} optical model with a {\it deep} potential
without a repulsive  core,     for  $\alpha$+$^{16}$O in Refs.\cite{Ohkubo1977,Michel1983}
and  for $\alpha$+$^{40}$Ca in Refs.\cite{Michel1977,Delbar1978}. 
 Furthermore the  clear observation of the Airy minimum of the nuclear rainbow in 
  $^{16}$O+$^{16}$O  scattering  at  $E_L$=350 MeV 
  \cite{Stiliaris1989}  showed that the  potential is {\it deep} 
in the  internal region \cite{Khoa2007}. 
The deep potentials in the internal region from  the ALAS and  rainbow are  inconsistent 
with the  repulsive core picture   concluded from the microscopic studies. 

\par
On the other hand, the deep potentials  are found to be  similar to a double folding model
 potential derived from  an effective two-body force.
 One might thus naively understand that the deep potentials,   hence the ALAS and
 nuclear rainbow phenomena, may be 
   a consequence of the  strong attractive nature of the nuclear forces.
However, in contrast to  traditional understanding,
I will show that the deep potentials, and therefore also the emergence of a nuclear rainbow
 and nuclear clustering, are  
a direct consequence of the Pauli  principle.

\par 
The purpose of this paper is to show that the Pauli  principle between
 nuclei causes    a strong Luneburg lens like {\it universal structural Pauli attraction}
 in the internal region, 
 which is in contrast to the traditional understanding  that the Pauli principle
 causes a repulsive core at short distances. The Luneburg lens like universal Pauli
 attraction  allows the  emergence  of the simultaneous existence of 
 cluster structures  near 
 the threshold energy in the low excitation energy region  and a nuclear rainbow
 in the high energy region.

\begin{figure}[t]
\includegraphics[keepaspectratio,width=8.6cm] {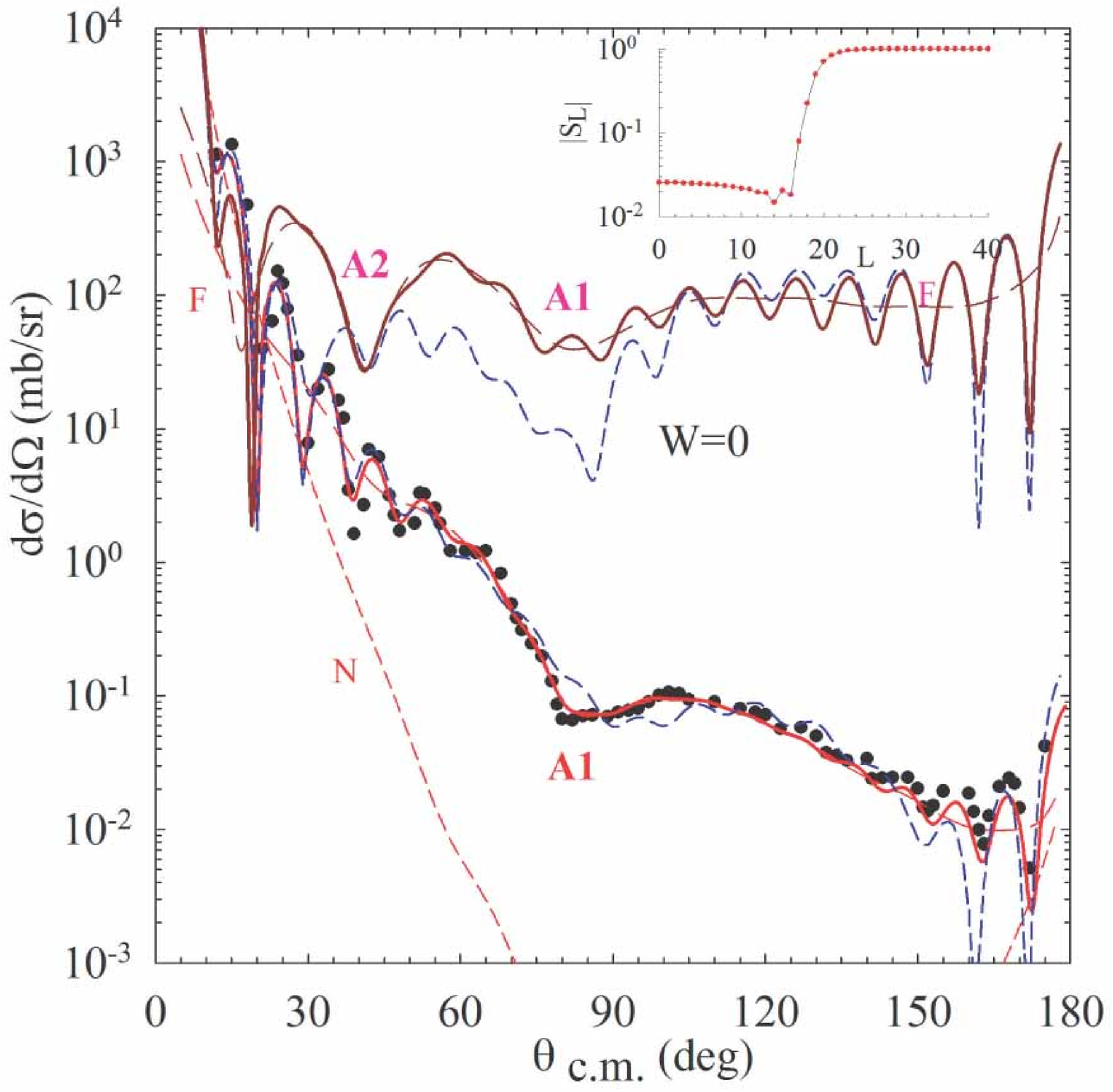}
 \protect\caption{\label{fig.1} {(Color online) 
 The experimental angular distribution (points) \cite{Delbar1978}  in 
  $\alpha$+$^{40}$Ca rainbow scattering at $E_L$=61 MeV and the  calculated one
  (red thick solid line), which is decomposed into 
the farside (red long dashed line)  and nearside (red medium dashed line) contributions.
The moduli  of $S$-matrices, $|S_L|$, in the inset (red filled circles) are connected by   lines  to guide the eye.
The angular distribution (brown thin solid line) and its farside component (brown long dashed line)
  calculated by switching off the imaginary potential (W=0) are also displayed. 
The cut-off calculations of $L=0-11$ partial waves with and without $W$ are shown by 
the blue short dashed lines.  
}
}
\end{figure}

\par 
In composite particle  scattering absorption is mostly strong,
 which makes it difficult to determine
the  potential up to the internal region without ambiguity. However there are
 some exceptions where absorption is weak or incomplete and the nuclear rainbow and  ALAS,
 in which scattering waves penetrate deep into the internal region, are observed in elastic
 scattering.
 $\alpha$+$^{16}$O \cite{Ohkubo1977,Michel1983,Michel1998} and $\alpha$+$^{40}$Ca 
\cite{Michel1977,Delbar1978,Michel1986,Ohkubo1988,Michel1998}  scattering are such typical
 examples,  for which a global  potential, which works over a wide range of energies,
has been determined. 

\begin{table*}[tbh]
\begin{center}
\caption{ \label{Table I}
 The overlap of the calculated wave functions of the $L=0$
 bound states  for $N <N_0$  with  
 the redundant Pauli forbidden  HO wave functions with the oscillator parameters
$\nu$=0.535, 0.32, 0.284 and 0.292
fm$^{-2}$ for  $\alpha$+$\alpha$, $\alpha$+$^{16}$O, $\alpha$+$^{40}$Ca  and
 $^{16}$O+$^{16}$O, respectively.    $\nu=m\omega/\hbar$
 with $m$ being the nucleon mass.
   $- V_0$ is the strength of the combined nuclear and  Coulomb  potential 
 near  the origin, $r=$0.01 fm in Fig. 2,  3(b),  4 and  5.
}
\begin{tabular}{ccccccccccccccc}
 \hline
  \hline
system &   $N_0$  & $V_0$ (MeV) & $N=0$ &$N=2$ & $N=4$ & $N=6$
 & $N=8$ &  $N=10$ &   $N=12$ &$N=14$   &$N=16$ &$N=18$&$N=20$ &$N=22$   \\
 \hline
 \hline
 $\alpha$+$\alpha$ &  4     &  119 &0.99&0.99&&&&&&&&&&  \\
 $\alpha$+$^{16}$O  &   8     & 134  &0.99&0.98&0.99&0.98&&&&&&&&   \\
 $\alpha$+$^{40}$Ca &  12     &  151 &1.00&1.00&1.00&1.00&0.99&0.93&&&&&&  \\
 $^{16}$O+$^{16}$O &  24   & 321 &0.98&0.93&0.91&0.89&0.89&0.91&0.93&0.96&0.96&0.90&0.74& 0.43 \\
 \hline                          
 \hline                          
\end{tabular}
\end{center}
\label{Table1}
\end{table*}

\par 
 In Fig.~1 the angular distribution in $\alpha$+$^{40}$Ca scattering  at $E_L$=61 MeV, 
 calculated using a   global potential with a Woods-Saxon squared form factor 
(thick solid line), which  works  well over a wide range of energies $E_L$=24 - 166 MeV \cite{Delbar1978},
is  compared with the  experimental data. 
The decomposition of the calculated cross sections into the farside
and nearside components shows that the minimum at  around 
 $\theta=80^\circ$ is caused by  farside refractive scattering and 
 is the first order Airy minimum $A1$ of the 
nuclear rainbow. The global potential can be   uniquely determined by   reproducing
  the Airy structure of the nuclear rainbow. 
The calculations in which the imaginary potential is switched off ($W=0$) show that 
 the minimum at around  $\theta=40^\circ$
 is a remnant of the Airy minimum $A2$, and the  broad bump  in the experimental 
 angular distribution in the $\theta=40-80^\circ$ region   is a remnant of the $A2$ Airy 
maximum. The moduli  of the $S$-matrices, $|S_L|$, which for $L$=0-11 is of the order 
of $10^{-2}$  (inset), shows
 that absorption is relatively weak. This makes   the observation of  a nuclear
 rainbow possible.
The cut-off calculations for the smaller $L$ values, $L$=0-11, (short dashed line) 
show that these partial waves 
 contribute to the correct  description of the   Airy structure, 
which is   also confirmed   in  the same  calculations with $W=0$ (short dashed line).
The global potential also reproduces well  the ALAS \cite{Delbar1978}, the
 $\alpha$+$^{40}$Ca fusion oscillations   in the  lower energy region 
$E_L$=10-27 MeV  \cite{Michel1986B} and the $\alpha$ cluster structure in $^{44}$Ti 
including the energy levels, $B(E2)$ values and
  $\alpha$ spectroscopic  factors \cite{Michel1986,Michel1998,Yamaya1990,Yamaya1998}.
  The semi-microscopic double folding potentials derived 
 from the effective  two-body Hasegawa-Nagata-Yamamoto (HNY) force \cite{Hasegawa1971}
 and the  density-dependent M3Y (DDM3Y) force \cite{Kobos1982} are similar to the global
 potential and    describe    the $\alpha$+$^{40}$Ca system over a wide range of energies as well
 \cite{Ohkubo1988,Atzrott1996}.

\begin{figure}[t]
\includegraphics[keepaspectratio,width=5.5cm] {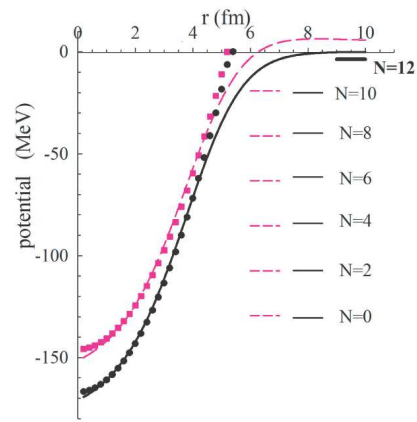}
 \protect\caption{\label{fig.2} {(Color online) 
 The global nuclear  potential (black solid lines) and the corresponding  Luneburg
 lens potential (black circles)  for the   $\alpha$+$^{40}$Ca  system. 
The potential  including the Coulomb potentials and corresponding  Luneburg lens potential 
are indicated by  long dashed lines (pink)  and  squares (pink), respectively.  
 The calculated   eigenstates for $L=0$  with  $N<N_0=12$ 
together with the $N=N_0$ band head  $0^+$ of  the 
   $\alpha$+$^{40}$Ca cluster state in $^{44}$Ti  are indicated by  horizontal solid lines.
The eigenenergies of the Luneburg potential including the Coulomb potential
 are indicated  by  horizontal dashed lines. 
}
}
\end{figure}

\par 
In Fig.~2 the global potential (solid lines) used in Ref.\cite{Michel1986}, D180 with 
the  potential  strength -180 MeV,  and the  potential including the Coulomb potential 
 (long dashed lines) are displayed.
 The internal region of  these  potentials resemble  the 
truncated harmonic oscillator (HO) potential well, which is called  a Luneburg lens potential, 
as displayed by the filled circles (black) and squares
 (pink), respectively. The  depth -$V_0$ and the truncation radius $R$, at which
the HO potential is zero,   are $V_0$=167 MeV and $R$=5.3 fm  
for the nuclear potential only  and $V_0$=146 MeV and $R$=5.2 fm
 for the combined  nuclear and  Coulomb potential. 
In Fig.~2 the states with $N<N_0=12$ are  Pauli forbidden,  where
   $N\equiv 2n+L$ with  $n$ being  the number of the nodes
in the wave functions.  The $N=12$ state   corresponds to the ground 
state  with  the $\alpha$+$^{40}$Ca cluster structure in $^{44}$Ti. 
The eigenenergies of the truncated HO potential (horizontal dashed lines) correspond well to
those of  the global potential.
In Table I  the overlap of the calculated wave functions with $N<N_0$ 
   in the   global potentials  with the redundant Pauli forbidden  HO  wave functions 
is almost complete.
This means that  the redundant Pauli forbidden  states
 of the RGM equations are  almost completely  embedded  in the  global potential.
The situation is almost the same for  the other $L<N_0 -2n$.

\begin{figure}[t]
\includegraphics[keepaspectratio,width=7.5cm] {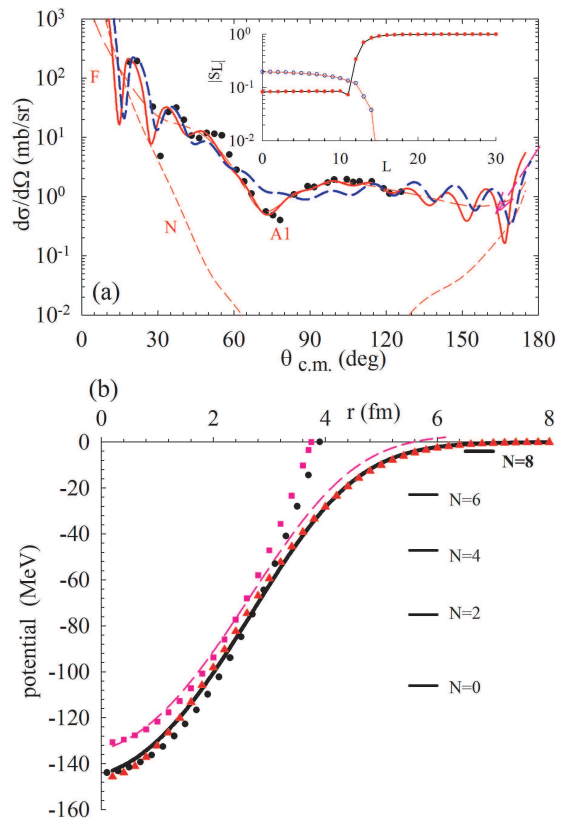}
 \protect\caption{\label{fig.3} {(Color online) 
 Same as Fig.~1 for panel (a) and Fig.~2 for panel (b) but for the $\alpha$+$^{16}$O system
 with $N_0=8$.
In the inset of panel (a) $|S_L|$ of the internal waves  are additionally 
 indicated by  unfilled circles. In panel (b)
  the  triangles (red) are the double folding  potentials derived 
from the  HNY  force   with -538 MeV for the triplet even state 
in the intermediate range \cite{Hasegawa1971}.
The $N=8$ (solid horizontal line) shows the  $\alpha$+$^{16}$O cluster ground 
 state in $^{20}$Ne.
}
}
\end{figure}

\par
This is  true for  other systems.
In Fig.~3(a) the experimental  angular distribution in $\alpha$+$^{16}$O rainbow scattering 
 at  $E_L$=49.5 MeV   is compared with the one calculated using  the phenomenological
 global potential, which was  determined from the 
systematic analysis of the  ALAS and nuclear rainbow scattering \cite{Michel1983}.
The  Airy structure      with the Airy minimum $A1$
 at around  $\theta=75^\circ$  followed by the broad
 Airy maximum $A1$ is well reproduced by the global potential.
The Airy  structure  is brought about by the farside refractive scattering. 
Furthermore, using the technique in Refs.\cite{Albinski82,Michel2001}, 
the Airy structure  is found to be caused by  the interference
 between  the two sub-amplitudes, i.e., the farside-subcomponent of the internal-waves, 
 which penetrate the potential barrier at
 the surface into the internal region,  and the farside-subcomponent of the barrier waves,
 which are reflected at the barrier. 
In fact,  in the inset of Fig.~3(a)  the moduli of
 the  $S$-matrix of the internal waves calculated using the technique in  Ref.\cite{Albinski82}
 are significantly large.  If one cuts off the
 contributions of the  partial  waves for $L=$0-7 (blue medium-dashed lines),  
 the Airy minimum is  destroyed  in disagreement with  the  experiment.
 This means the    waves with smaller $L$ values
  contribute to the   correct reproduction of the  Airy structure, i.e.,  
  to constraining the shape and depth of the potential in the internal region. 
In Fig.~3(b) the global potential for the $\alpha$+$^{16}$O system
 with the energy-dependent parameter $\alpha$=3.02 in Ref.\cite{Michel1983}  is shown.
This potential, which reproduces well the observed $\alpha$ cluster structure in $^{20}$Ne, 
 the energy levels, $B(E2)$ values and
   $\alpha$ widths \cite{Michel1983,Michel1998}, resembles  the semi-microscopic
 double folding potentials 
 derived  from the  HNY force  (triangles in Fig.~3) \cite{Michel1983,Ohkubo1977}
 and the DDM3Y force  \cite{Abele1993,Hirabayashi2013} well.
The internal region of the global  
potential (solid line) is well simulated by the Luneburg lens truncated HO potential with 
$V_0$=131 MeV and $R$=3.75 fm (filled circles) 
and,  when the Coulomb potential is added (long dashed line),  
 with $V_0$=144 MeV and  $R$=3.9 fm (filled squares).
The $N=8$ state corresponds to the  ground state with the 
$\alpha$+$^{16}$O cluster structure in $^{20}$Ne. 
 Table I shows that  the overlap of the wavefunctions of the  states with $N<$8  
  and   the redundant Pauli forbidden HO wave functions is almost complete.

\begin{figure}[t]
\includegraphics[keepaspectratio,width=5.5cm] {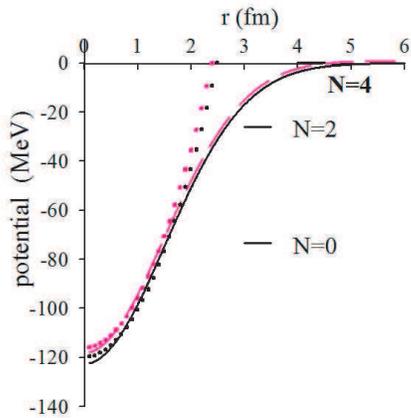}
 \protect\caption{\label{fig.4} {(Color online) 
 Same as Fig.~2 but for the $\alpha$+$\alpha$ system with $N_0=4$.
The solid horizontal line shows the $N=4$  ground  state with 
 $\alpha$+$\alpha$ cluster structure  in $^{8}$Be.
}
}
\end{figure}

\begin{figure}[t]
   \includegraphics[width=5.5cm]{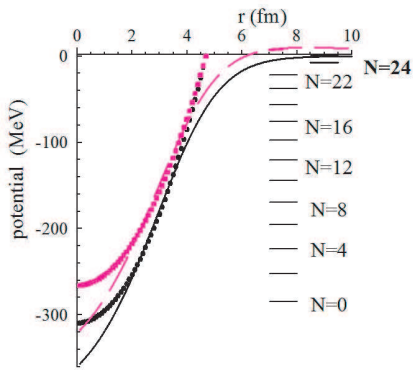}
 \protect\caption{(Color online)
 Same as Fig.~2 but for the  $^{16}$O+$^{16}$O  system with $N_0=24$.
The solid horizontal line shows the $N=24$ $0^+$ state with  
$^{16}$O+$^{16}$O  cluster  structure in $^{32}$S.
  \label{fig5}
}
\end{figure}

\par 
 In Fig.~4 the global potential for  $\alpha$+$\alpha$ \cite{Buck1977}, which reproduces 
 the experimental phase shifts of elastic scattering over a wide range of energies, 
is displayed.
The internal region of the potential without (with) the Coulomb potential is well 
simulated  by the  truncated HO potential  with    $V_0$=120 MeV and $R$=2.5 fm 
(116 MeV and 2.4 fm).
In Fig.~5 the  global potential for  $^{16}$O+$^{16}$O  in Ref.\cite{Ohkubo2002}  
 is displayed.
The global potential reproduces the rainbow scattering 
\cite{Stiliaris1989,Khoa2000,Nicoli1999,Michel2001,Khoa2007} 
and molecular structure with the  $^{16}$O+$^{16}$O cluster structure 
in a unified way \cite{Ohkubo2002}.
  The internal region of the potentials 
 resemble   the Luneburg lens  truncated HO  potentials well, with 
$V_0$=266 MeV and $R$=4.7 fm when the Coulomb potential is included 
  and with  310 MeV and 4.8 fm for the nuclear potential only.
For the $^{16}$O+$^{16}$O system it is noted that the  region $r<2$ fm has some ambiguity
and the  slightly different folding potential   in this region can   reproduce 
the $^{16}$O+$^{16}$O scattering equally well \cite{Nicoli1999}.
In Table I the  overlap of the  states for $N < N_0$  with the redundant Pauli forbidden 
 HO wave functions is  large except  for $N=22$  near the threshold energy.

\par
Thus the physical wave functions  with $N \geq  N_0$ generated by the  global  potentials
 are  found to be   orthogonal  to the redundant Pauli forbidden HO wave functions  in the RGM.
This orthogonality is closely related   to the
shape and  depth of the potential in the  internal region, i.e., the Luneburg lens like
 truncated HO potential.
  I will now show theoretically that   the global potentials have a Luneburg
 lens like universal Pauli attraction   in the  internal region. 
 The RGM equation for the antisymmetrized wave function for two clusters 
 that are assumed to have   HO shell model  wave functions with the  size parameter 
 $\nu$ and  spin 0  is given by 
\begin{eqnarray}
 (T_r+V_D(r)-E)\chi_{\it L}({\it r}) + \int K(r,r^\prime)dr^\prime
\chi_{\it L}({\it r}^\prime)
 & = & 0
\label{eq:eq4}
\end{eqnarray}
\noindent where   $\chi_{\it L}({\it r})$, $T_r$, $V_D(r)$, $E$ and $K(r,r^\prime)$  
are the relative wave function,   kinetic energy operator of the relative
 motion,    direct (double folding) potential, relative energy  and   exchange kernel,
respectively. Since one knows that the local potential  works very well, if one  approximate
 $K(r,r^\prime) = V_{P}(r) \delta(r-r^\prime)$, Eq.~(\ref{eq:eq4}) becomes
a local potential equation 
 \begin{eqnarray}
 \{T_r + V_D(r) + V_{P}(r)-E\}\chi_{\it L}({\it r}) & = & 0,
\label{eq:eq7}
\end{eqnarray}
\noindent with  the local potential $V(r)\equiv V_D(r) + V_{P}(r)$.
One can impose  the   eigenfunctions   
$\chi_{\it L}^{(n)}({\it r})$ with 
 $n< (N_0 -L)/2$ to   satisfy
 \begin{eqnarray}
 \{T_r +  V_{HO}(r)-(2n+L+3/2)\hbar \omega\} \chi_{\it L}^{(n)}({\it r})  & = & 0,
\label{eq:eq8}
\end{eqnarray}
\noindent where $V_{HO}(r)$ is the HO potential with
a depth  $- V_0$ at $r=0$ and  the size parameter $\nu$.  
This guarantees that the  physical  wave functions of Eq.~(\ref{eq:eq7}) with 
$N\ge N_0$  are orthogonal to the redundant Pauli forbidden  states. 
This is satisfied when $V(r)=V_{HO}(r) $
in the internal region $r < R$,  where $R$ is the size of the Luneburg lens, which is a HO 
  potential  truncated at $r=R$  as given below. 
Thus in order that the wave functions of the physical states  satisfy the Pauli principle,
the local potential should resemble a deep HO potential  in the internal region,
i.e., a Luneburg lens potential. 
 When   the  $V_D(r)$ itself  resembles 
 a deep HO  as seen for the HNY force (triangles)  in Fig.~3(b),    
 the $V_{P}(r)$ is  small \cite{Ohkubo1977}. 
On the other hand, when   $V_D(r)$ is  repulsive (in the case of,
 for example,    Brink-Boeker force B1),   the  $V_{P}(r)$ must be    deep so that
  the  $V(r)$  resembles  a Luneburg lens potential. 
 Thus    the Pauli principle plays the role  generating a  $ V_{P}(r)$ so that 
 the   $V(r)$  resembles  a  Luneburg lens like HO potential in the internal  region.

\par
A Luneburg lens with a radius $R$   is a lens that refracts all the parallel
 incident trajectories to the focus $R_f$  ($< R$). For such a lens the 
 refractive index $n$ is given by 
\begin{equation}
n^2(r \leq R) =  ({R_f^2-r^2+R^2})/{R_f^2}, \quad    n(r > R) = 1.
\end{equation}
The potential having this property \cite{Michel2002} is 
\begin{equation}
V(r \leq R) = V_0 \left(  {r^2}/{R^2}-1 \right),\quad V(r > R) = 0,
\end{equation}
\noindent where $V_0 = E (R/R_f)^2$ is the depth at $r=0$. 
\noindent This is   a HO potential truncated at $r=R$.
 The outer region  of the nuclear potential has a diffuse surface, and so deviates from 
the ideal Luneburg lens. This causes   astigmatism to occur, which
 is nothing but the emergence of a  nuclear rainbow.
Thus  the emergence of the nuclear rainbow is
due  the properties of both  the Luneburg lens like potential in the
internal region  and the diffuse attraction in the outer region.
The values of the strength of the potential in Table I  are consistent with those    
 evaluated  from the constraint of the  Pauli  principle at $E$=0, i.e., 
$V_0 = (N_0+3/2)\hbar \omega$,
which are 121, 125, 157 and 305 MeV for  $\alpha$+$\alpha$, $\alpha$+$^{16}$O,
  $\alpha$+$^{40}$Ca and $^{16}$O+$^{16}$O,  respectively.
Thus the deep nature of the potential is a direct consequence of the Pauli principle.
This  explains why a rainbow occurs  in nuclear scattering in the 
potentials that generate cluster states at lower energies, near the threshold energy.
A deep  double folding  potential   derived from  a density-dependent effective two-body
 force, such as the DDM3Y force \cite{Kobos1982}, resembles the Luneburg lens like potential
and  has been  successfully used in scattering  and structure studies 
\cite{Khoa2007,Atzrott1996,Abele1993,Hirabayashi2013,Khoa2000,Nicoli1999}. 
 According to  the present study, for which  the depth  and shape in the internal region are
determined to be  constrained  by the Pauli principle,
it seems that the Pauli principle manifests itself through the density-dependence 
 as well as  the exchange terms \cite{Gupta1984,Chaudhuri1985,Khoa1988}.

\par
From the structure viewpoint,   shell model wave functions in  the  HO potential,
 which have almost complete overlaps with the Pauli forbidden states embedded in the local
 potential as shown  in Table I, 
are equivalent to the cluster representation of
 Wildermuth  and Kanellopoulos \cite{Wildermuth1958,Wildermuth1966,Bayman1958}
 and can be represented  by  the $SU(3)$ model
 \cite{Elliott1958}. This wave function with a Gaussian tail is damped at the surface.
   On the other hand, in the present   local potential cluster model, which has an 
attractive potential  with an exponential tail at the surface similar to  a  Woods-Saxon
 potential,  the inner oscillations of the wave function are damped
\cite{Tamagaki1965,Tamagaki1968,Hiura1972} due to the orthogonality to the Pauli forbidden 
states with $N<N_0$  embedded in the Luneburg lens like  potential. This   brings
 about  the enhancement  of the amplitude  of  the wave  function  at the surface, 
i.e., emergence and development of cluster structure.  
  Thus the Pauli principle plays the dual role of
  causing (i) the  {\it shell model} potential with a deep HO shape as the  structural Pauli 
attraction and (ii) the {\it cluster structure} with the
   damped inner oscillations and    enhanced surface amplitude   by the
 orthogonality to the embedded Pauli forbidden states due to (i).

\par
When two  nuclei with a typical shell model structure such as the double magic nucleus
 $\alpha$ particle, $^{16}$O, $^{40}$Ca and $^{208}$Pb  come closer, the universal
Pauli attraction inevitably makes possible the emergence of a  cluster
structure slightly above the highest Pauli-forbidden state because of the diffuse surface,
 i.e.,  near the threshold energy of the compound system. 
This is the reason why the $\alpha$ cluster structure  typically appears 
 for the double magic core + $\alpha$  systems like $^8$Be, $^{20}$Ne, $^{44}$Ti and $^{212}$Po. 
 The shell model 
  structure of the internal constituent nuclei (dynamical nature) and the existence of
  redundant Pauli forbidden states due to the  Pauli principle in the wave functions of 
 the relative motion  (kinematical nature)  are closely interrelated for  the emergence of
 both  the  cluster structures in the compound system and  nuclear rainbows.
The Pauli principle does not only provide  
the {\it raison d'\^etre} for 
the  shell structure  of nuclei but also for the emergence of the  cluster structure 
near the threshold.
This will  not be limited to closed nuclei and  two nuclear systems as long as 
redundant Pauli forbidden states exist.
Also, the nucleon-nucleon potential  may  have a strong Pauli attraction 
due to the Pauli principle rather than the   repulsive core
 \cite{Otsuki1964,Otsuki1965,Aoki2012},
 the effects of which  could be seen in few body systems.

\par
To summarize,     the existence of a Luneburg lens like universal
 structural Pauli attraction  in the  internal region of  nucleus-nucleus interaction 
has been    shown.
 This  is different from the  traditional view that a structural repulsive  core exists
 at short distances.
It is  found that the depth and the shape of the potential in the internal region is
 constrained to a Luneburg lens  like truncated harmonic oscillator potential  by the
Pauli principle. 
  The present work reinforces the empirical  threshold  rule,  which had intuitively been 
 understood  to be due
 to the   saturation property  of the nuclear force. 
   The  emergence of a cluster state near the threshold energy can now be seen as 
     a consequence of both  the Pauli principle in the internal region
and  diffuse attraction in the outer region. 
 The emergence of cluster structures and  rainbows are unified as a consequence of
 a Luneburg lens like  universal structural Pauli attraction in the internal region 
(kinematical) and  diffuse attraction in the outer region (dynamical).

 \par
The author thanks the Yukawa Institute for Theoretical Physics, Kyoto University for
 the hospitality extended  during a stay in February 2016 where this work 
 was completed.

\end{document}